\documentclass{article}

\newcommand{\msun}{{\rm M}_\odot}

\newcommand{\rds}{\dot{\rho}_\ast}
\newcommand{\rdss}{\dot{\rho}_{\ast,{\rm SN}}}
\newcommand{\rdsa}{\dot{\rho}_{\ast,{\rm AGN}}}
\newcommand{\rdsb}{\dot{\rho}_{\ast,{\rm SN+AGN}}}

\usepackage{natbib}
\usepackage{graphicx}

\usepackage[margin=1.5in]{geometry}

\begin{document}
\title{The interaction between feedback from active galactic nuclei and supernovae}
\author{
        C. M. Booth$^{*}$ \\
        Department of Astronomy \& Astrophysics, The University of Chicago,\\
        Chicago, IL 60637
            \and
        Joop Schaye\\
        Leiden Observatory, Leiden University, P.O. Box 9513, 2300 RA Leiden,\\
        The Netherlands\\
          \\
        $^{*}$Correspondence to cmbooth@oddjob.uchicago.edu
}
\date{}
\maketitle

\begin{abstract}
Energetic feedback from supernovae (SNe) and from active galactic nuclei (AGN) are both important processes that are thought to control how much gas is able to condense into galaxies and form stars.  We show that although both AGN and SNe suppress star formation, they mutually weaken one another's effect by up to an order of magnitude in haloes in the mass range for which both feedback processes are efficient ($10^{11.25}\,\msun<m_{200}<10^{12.5}\,\msun$).  These results demonstrate the importance of the simultaneous, non-independent inclusion of these two processes in models of galaxy formation to estimate the total feedback strength.  These results are of particular relevance to the interpretation of results from hydrodynamical simulations that model only one of the feedback processes, and also to those semi-analytic models that implicitly assume the effects of the two feedback processes to be independent.
\end{abstract}


\section*{Introduction} \label{sec:intro}
Feedback from the formation of massive stars, probably in the form of large-scale winds driven by supernova (SN) explosions, and from accreting black holes (BHs) associated with active galactic nuclei (AGN), are thought to regulate the star formation rates (SFRs) of low- and high-mass galaxies, respectively (for a review see e.g. \citep{bens10}). Observationally the two feedback effects are frequently seen to be coeval.  In a global sense, the evolution of the cosmic SFR and the luminosity density of quasars are tightly correlated \citep{boyl98} and on the scale of individual objects post-starburst galaxies are found to preferentially host active BHs \citep{kauf03}.  Both AGN and SN feedback are now routinely included in semi-analytic and numerical simulations.  However, it is frequently assumed, particularly in some semi-analytic models \citep[e.g.][]{bens03,delu04,bowe06,guo11}, that the efficiency of one form of feedback is unperturbed by the inclusion of the other.  

In semi-analytic models it is generally assumed that the amount of gas reheated by SNe is $\propto \dot{m}_{\ast}v^{\alpha}$, where $\dot{m}_{\ast}$ is the galaxy star formation rate, $v$ is some characteristic velocity associated with the galaxy and $\alpha$ is a free parameter.  AGN feedback is included in most semi-analytic models by coupling the BH directly to the halo without regard for the galaxy, either by assuming that the BH growth decreases the cooling rate in the hot halo in proportion to the BH accretion rate \citep{guo11}, or by shutting off cooling if the BH is massive enough for the Eddington luminosity to exceed some multiple of the cooling luminosity of the halo \citep{bowe06}.  In some models \citep{bowe08,bowe12} the effect of AGN feedback is not limited to halting the cooling of the halo, but can remove gas entirely from the halo, which improves model descriptions of the hot halo gas that surrounds galaxies.

It is by no means clear that SN and AGN feedback act independently of each other. Both feedback processes redistribute gas inside of galaxies in a complex and non-linear way.  It is possible that the effect of one feedback process, e.g.\ the factor by which the cooling of hot halo gas is reduced or the amount of cold gas that is heated or ejected, depends on the presence of other feedback processes. Indeed, it has been demonstrated that SN feedback and photo-heating, which semi-analytic models assume to act independently, amplify each other's suppression of the SFR of low-mass galaxies \citep{pawl09,finl11}.

The aim of this paper is to use self-consistent hydrodynamical simulations to investigate the mutual amplification of AGN and SN feedback.  We will show that in haloes in the mass range $10^{11.25}\,\msun<m_{\rm 200}<10^{12.5}\,\msun$ AGN and SN feedback act to \emph{suppress} one another's effects by up to an order of magnitude.

\section*{Method}
\label{sec:method}

This study is based on smoothed particle hydrodynamics (SPH) simulations of representative volumes of the Universe.  Gravitational forces and the equations of hydrodynamics are solved using a significantly extended version of the parallel PMTree-SPH code {\sc gadget iii} \citep[last described in ][]{spri05b}, a Lagrangian code used to calculate forces on a particle by particle basis.  All of our simulations assume a $\Lambda$CDM cosmology with parameters determined from the 3-yr Wilkinson Microwave Anisotropy Probe (WMAP) results, $\Omega_{\rm m} = 0.238$, $\Omega_\Lambda=0.762$, $\Omega_{\rm b} = 0.0418$, $h$ = 0.73, $\sigma_8 = 0.74$ and $n_{\rm s} = 0.951$. These values are consistent with the 7-yr WMAP data \citep{koma11} except that the parameter $\sigma_8$ is 2$\sigma$ lower in the WMAP 3-year data than allowed by the WMAP 7-year data.

In addition to treating gravitational and hydrodynamic forces, the simulations need to follow the galaxy formation processes that happen on small scales.  The simulations track star formation, SN feedback, BH growth and AGN feedback,  radiative cooling and chemodynamics, as described in \citep{scha08,dall08,boot09,wier08,wier09}, respectively. This physical model is denoted \lq\emph{AGN}\rq\, in the OWLS suite of simulations \citep{scha10}. 

For the purposes of this work, our prescriptions for SN and AGN feedback are the most important aspects of the physical model so we describe these in some detail here.  Feedback from SNe is implemented by injecting approximately 40\% of the energy released by Type II SNe locally as kinetic energy. The rest of the energy is assumed to be lost radiatively. Each newly formed star particle kicks on average 2 of its neighbouring gas particles into the wind. The initial wind velocity is 600 km/s, which is consistent with observations of galaxy outflows \citep[e.g.][]{veil05}.

BH growth and AGN feedback is implemented using the method of \citep{boot09} which is, in turn, a modification of that from \citep{spri05}. We regularly run a friends-of-friend halo finder and insert seed mass BHs ($m_{\rm seed}=10^5\,\msun$) into every halo of mass $>10^{10}\,\msun$ that does not yet contain a BH.  These seed BHs then grow both through merging and gas accretion (which is limited to the Eddington rate). Accretion rates in low-density gas ($n_{\rm H}<10^{-1}\,$cm$^{-3}$) are assumed to be equal to the Bondi-Hoyle rate.  For higher-density, star-forming gas the Bondi-Hoyle rate is boosted by a factor $(n_{\rm H}/10^{-1}\,{\rm cm}^{−3})^2$ to compensate for the lack of a cold, interstellar gas phase and the finite resolution (see \citep{boot09} for a full discussion). The BH growth rate is related to its accretion rate by  $\dot{m}_{\rm BH} = (1-\epsilon_{\rm r})\dot{m}_{\rm accr}$, where $\epsilon_{\rm r}=0.1$ is the radiative efficiency of the BH.  One significant source of uncertainty in simulations of AGN feedback is in the description of BH accretion rates \citep{hopk11,hobb12}, but this uncertainty does not significantly affect the results presented here.  We showed previously \citep{boot10} that the masses of simulated BHs are set by self-regulation.  BHs grow until they are capable of injecting energy at a sufficiently high rate to counteract gas inflow.  This means that in the simulations, although the instantaneous accretion rate depends on the accretion model, the final mass of massive BHs is insensitive to the accretion model provided the accretion rate would become high in the absence of AGN feedback \citep{boot09}.

The amount of energy available for AGN feedback is then given by $\dot{E}=\epsilon_{\rm f}\epsilon_{\rm r}\dot{m}_{\rm accr}c^2$, where $c$ is the speed of light and $\epsilon_{\rm f}$ is a free parameter, the \lq feedback efficiency\rq.  The BH builds up a reservoir of energy until it is capable of heating one gas particle by a temperature $\Delta T=10^8$~K.  This energy is then injected thermally into one of the BH's neighbours, chosen at random.  The feedback efficiency is tuned to reproduce the global BH density at $z=0$.  In the fiducial runs $\epsilon_{\rm f}=0.15$.   

We note that the precise values for parameter choices such as energy efficiencies do not reflect in detail the physical processes that are happening on small scales.  This is because in cosmological simulations, limited resolution means that the effects of the feedback parameters are dependent upon other sub-grid models, such as the numerical equation of state for star-forming gas.  Our approach to calibrating these parameters, therefore, is simply to constrain them to be physically plausible, and to tune them to match a particular observable.

The fiducial simulations analysed in this work are performed in cubic volumes of 50\,co-moving Mpc/$h$ and contain $256^3$ particles of both gas and dark matter (DM).  For the assumed cosmological parameters this corresponds to DM and (initial) gas particle masses of $4.1\times 10^8\,\msun/h$ and $8.7\times10^7\msun/h$, respectively.  Initial conditions are generated with {\sc CMBFAST} \citep{selj96}, and evolved linearly to the simulation starting redshift of $z=127$.  The comoving gravitational force softening is set to 1/25 of the initial mean interparticle spacing and is limited to a maximum physical scale of 2~kpc/$h$, which is reached at $z\approx 3$.  

Each simulation is run four times, starting from the same initial conditions.  One realisation includes no feedback processes, the next two include either AGN or SN feedback. The final run includes both AGN and SN feedback. Simulations with this last physical model and the same resolution as used here reproduce the observed $z=0$ relations between BHs and the mass and velocity dispersion of their host galaxies \citep{boot09} as well as the observed optical and X-ray properties of the groups in which they reside \citep{mcca10}. 

In the simulation where we treat AGN feedback but not SN feedback, the AGN grow over-massive when compared to the observed global density of BHs \citep{shan04} and the $z=0$ relations between BH mass and galaxy bulge mass \citep[e.g.][]{hari04} and velocity dispersion \citep{trem02}. We therefore run one additional simulation in which the AGN efficiency is increased by a factor 6.87 (the ratio of the global BH densities in the simulations with and without SN feedback) to reproduce the scaling relations even in the absence of SN feedback.

For our analysis we use snapshots of the simulations, which are saved at discrete output redshifts with interval $\Delta z=0.125$ at $0\le z\le0.5$, $\Delta z=0.25$ at $0.5<z\le4$ and $\Delta z=0.5$ at $4<z\le6$. At each of these outputs we use a Friends-of-Friends halo finder with linking length $b=0.2$ to obtain a list of DM haloes.  Their masses are then determined using the {\sc SubFind} code \citep{dola09}, which employs a spherical-overdensity criterion centered on the most bound particle in each halo.  All halo masses quoted in this work are spherical overdensity masses defined as the mass within a sphere that encloses a mean density 200 times the critical density of the Universe. We limit our analysis to haloes of mass $m_{200}>10^{10.75}\,\msun$, corresponding to $>100$ DM particles.

\section*{Results}\label{sec:results}

\begin{figure}
\begin{center}
\includegraphics[width=\textwidth]{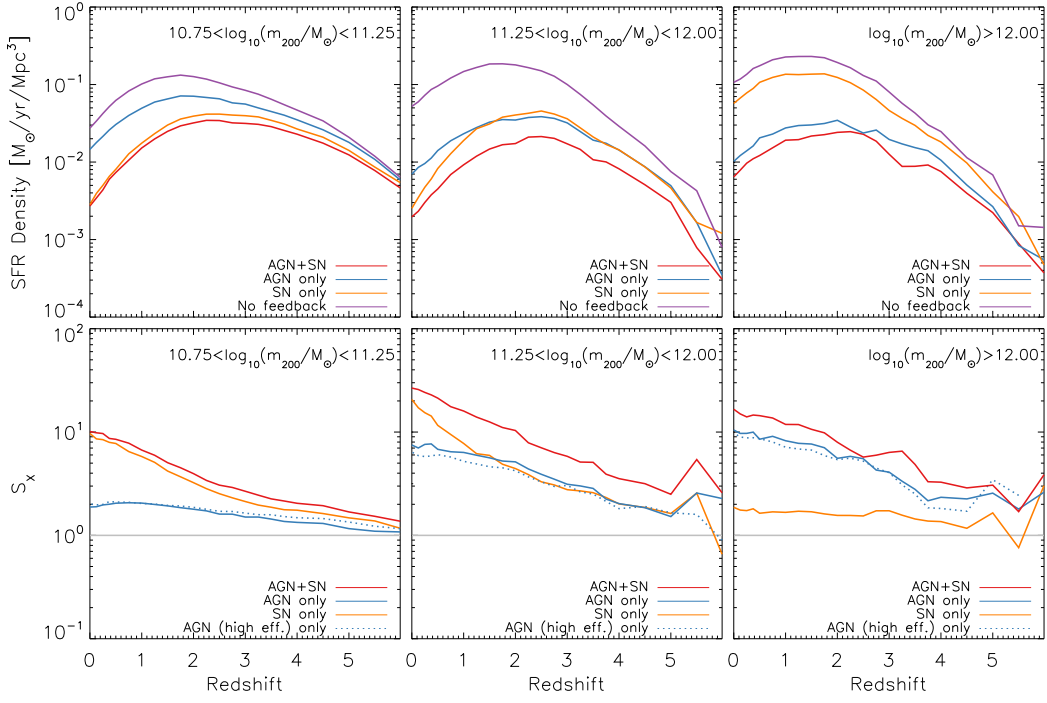}
\end{center}
\caption{\label{fig:sfr}  Effect of SN and AGN feedback on the evolution of the cosmic SFR density contributed by different halo masses. Different colours represent different feedback models. Purple curves show simulations that include no feedback; red curves include both AGN and SN feedback and orange (blue) curves show simulations that include only SN (AGN) feedback. \emph{Top panels:} the contribution to
the cosmic SFR density from haloes with different masses as a function of redshift. \emph{Bottom panels}: the factor by which the SFR is decreased relative to the simulation that includes no feedback.  The horizontal, grey line in each of these panels shows no-suppression.  The blue, dotted lines demonstrate that changing the AGN feedback efficiency has almost no effect on the overall SFRs. At the low-mass end ($m_{200}<10^{11.25}\,\msun$) SN feedback dominates and AGN feedback dominates at the high-mass end ($m_{200}>10^{12}\,\msun$), but at intermediate masses both feedback processes contribute to the decrease in the SFR.}
\end{figure}

We begin by considering the effect of each feedback process on the SFR of galaxies residing in haloes of a fixed mass.  Each panel in the top row of Fig.~\ref{fig:sfr} shows the contribution of haloes in a given mass range to the cosmic SFR as a function of redshift in each of the simulations. The colour denotes the physical model: no feedback (purple), SN feedback only (orange), AGN feedback only (blue) and both AGN and SN feedback (red).  We denote the global SFR in the simulation that includes no feedback with $\rds$, and the simulations that include only SN feedback, only AGN feedback and both forms of feedback as $\rdss$, $\rdsa$ and $\rdsb$, respectively.  We can then define suppression factors, $S_{\rm X}\equiv\frac{\rds}{\dot{\rho}_{\ast{\rm ,X}}}$, where $X$ represents one of the subscripts introduced above.  The bottom row of Fig.~\ref{fig:sfr} shows the suppression factors corresponding to each of the upper panels.  In each of the lower panels, the horizontal, grey line shows no-suppression (i.e. $S_{\rm X}=1$). 

In the least massive haloes ($m_{200}<10^{11.25}\,\msun$; left panels) AGN feedback has not yet become efficient so the simulations with no feedback (purple) and AGN feedback (blue) lie very close to one another. The simulations with SN feedback (orange) and both AGN and SN feedback (red) also nearly overlay one another.  This indicates that in this mass range, SN feedback dominates the suppression of the SFR.  Conversely, at the highest masses ($m_{200}>10^{12}\,\msun$; right panels) SN feedback becomes increasingly inefficient and now the simulations that include no feedback (purple, solid curve) and only SN feedback (orange) lie close to one another, while the AGN only (blue) and AGN and SN feedback (red) lines also lie very close to one another, indicating that in this mass range SN feedback does little to suppress the SFR.  At intermediate masses ($10^{11.25}\,\msun<m_{200}<10^{12}\,\msun$; middle panels) the two feedback processes suppress the SFR by comparable amounts, although at low redshift AGN feedback becomes less efficient. 

In the absence of SN feedback, the BH population grows to be much more
massive than observed.  We therefore consider one extra simulation in
which $\epsilon_{\rm f}$ is increased by a factor of 6.87 (the ratio
of the global BH densities in the simulations without and with SN
feedback) in order to bring the total mass in BHs back in line with
observations.  The dotted, blue curves in the bottom panels of
Fig.~\ref{fig:sfr} show suppression ratios in this simulation.  It is
notable that this line is almost identical to the AGN only simulation
with the standard efficiency, indicating that in both cases the AGN
injects the same amount of energy into its surroundings and suppresses
the SFR by the same factor.  This occurs because if the BH feedback
efficiency is increased by some factor, the total mass in BHs
decreases by nearly the same factor, so that the total amount of
energy output is almost independent of $\epsilon_{\rm f}$.  This
suggests that BHs are growing until they have output some critical
amount of energy, at which point they are capable of regulating their
own growth (see \citep{boot09,boot10} for a full discussion of this
point).  We therefore do not show the high-efficiency simulation in
the rest of this paper but note that our results are insensitive to
this choice and that the only effect of the feedback efficiency is to
scale the BH masses. In low mass haloes, where SN are able to self
regulate, the same relationships hold.  For example, it has been
demonstrated in the OWLS simulations that doubling the amount of
energy available from SN halves the instantaneous accretion rate onto
low mass galaxies\citep{vand11b}, which is also consistent with the
self-regulation argument presented here. In addition, it has been
shown in the same simulations \citep{haas13} that in low mass
galaxies, where AGN feedback is unimportant, the star formation rate
is inversely proportional to the amount of energy injected per unit
stellar mass formed. This implies that the rate at which energy is
injected is independent of the assumed efficiency of the feedback, as
would be expected if galaxies regulate their growth by generating
outflows that match the cosmological accretion.

\begin{figure}
\begin{center}
\includegraphics[width=8.3cm]{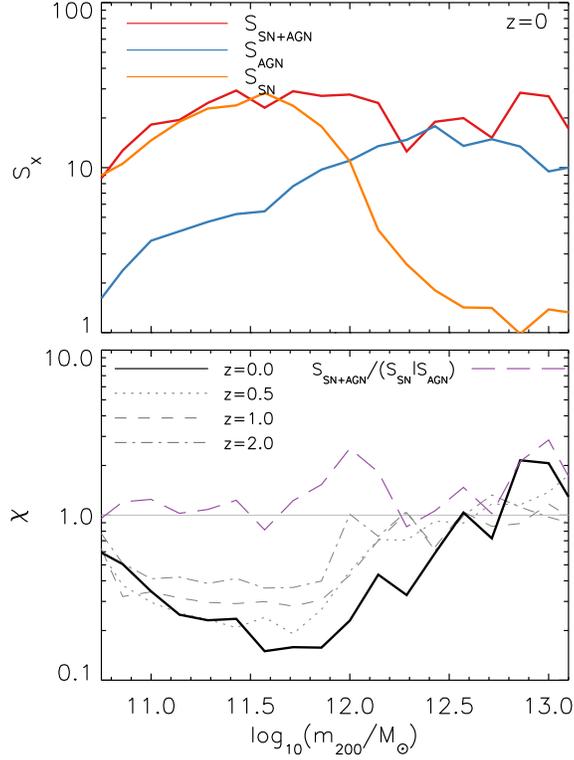}
\end{center}
\caption{\label{fig:chi} \emph{Top panel:} Present-day star formation suppression factors as a function of halo mass for the simulations that include SN feedback (orange), AGN feedback (blue) and both types of feedback (red).  \emph{Bottom panel:} Amplification factor, $\chi$ (Eq.~\ref{eq:chi}), as a function of halo mass.  Each curve shows results for a different redshift, as indicated in the legend.  Values of $\chi=1$ indicate that SN and AGN feedback operate independently of one another.  Values $\chi<1$ indicate that SN and AGN feedback mutually suppress one another's effects.  At low redshift, in haloes with masses $10^{11}-10^{12.5}\,\msun$, the two feedback effects weaken one another's impact on the SFR by almost an order of magnitude relative to the case in which they operate independently.  The purple, dashed line in the bottom panel of this figure shows the suppression factor obtained from using $S_{\rm SN+AGN}/S_{\rm SN}$ in haloes with $m_{200}<10^{12}\,\msun$ and $S_{\rm SN+AGN}/S_{\rm AGN}$ for more massive haloes.  At all halo masses, this quantity remains within a factor of two of unity, demonstrating that using a sharp cutoff where only SN (AGN) feedback operates in low mass (high mass) haloes provides a reasonable approximation to our results.}
\end{figure}

In the top panel of Fig.~\ref{fig:chi} we show the $z=0$ suppression factors as a function of halo mass for SN feedback (orange), AGN feedback (blue) and both feedback mechanisms (red).  SN feedback accounts for most of the suppression of the SFR in haloes with masses $\le 10^{12}\,\msun$, but its efficiency falls to almost zero for larger halo masses.  Above $m_{200}=10^{12}\,\msun$, AGN feedback accounts for the majority of the suppression of the SFR.

In order to quantify the mutual amplification of SN and AGN feedback, we define an \lq amplification factor\rq\, \citep[][]{pawl09} for the two feedback processes using their suppression ratios
\begin{equation}
\label{eq:chi}
\chi\equiv \frac{S_{\rm SN+AGN}}{S_{\rm SN}\times S_{\rm AGN}}\,.
\end{equation}
A value $\chi=1$ indicates that AGN and SN feedback suppress the SFR independently of one other. A value $\chi>1$ ($\chi<1$) indicates that AGN and SN feedback amplify (weaken) each other's ability to suppress the SFR relative to the case where they act independently.

The solid curve in the bottom panel of Fig.~\ref{fig:chi} shows the amplification factor, $\chi$, as a function of halo mass for $z=0$.  At both low ($m_{200}<10^{11}\,\msun$) and high ($m_{\rm 200}>10^{12.5}\,\msun$) halo masses, the mass ranges where only one of the two feedback processes are effective, the amplification factor $\chi\sim 1$, indicating that each process operates independently of the other.  However, in the intermediate mass range $\chi\ll 1$, indicating that \emph{the effect of including both AGN and SN feedback is almost an order of magnitude weaker than it would be if they each reduced the SFR by the same factor as when they act in isolation}.  In the lower panel of Fig.~\ref{fig:chi}, it appears that $\chi>1$ for high halo masses, but this is not statistically significant, and disappears at other redshifts, or if the data is binned differently.

The magnitude of this suppression is largest at redshift zero (solid, black curve), but the same effect exists up to high redshift (dotted curves).  The highest halo mass at which SN feedback is able to have an effect depends on when the winds become pressure confined by the interstellar medium \citep{dall08}.  At higher redshifts, ambient gas densities (and galaxy gas fractions) are higher and thus SN winds are only efficient in haloes of lower mass, and the range of halo masses over which both feedback processes are effective (and thus suppress one another) decreases.  The dashed, purple curve in the bottom panel of Fig.~\ref{fig:chi} shows the amplification factor for the full simulation relative to the SN-only case for $m_{200}<10^{12}\,\msun$ ($=S_{{\rm SN+AGN}}/S_{{\rm SN}}$) and relative to the AGN only case for $m_{200}>10^{\rm 12}\,\msun$ ($=S_{{\rm SN+AGN}}/S_{{\rm AGN}}$).  This curve lies close to unity for all halo masses, demonstrating that a fairly accurate approximation to our full results can be obtained if models assume a sharp transition between SN and AGN feedback at a halo mass of $10^{12}\,\msun$.

Eq.~\ref{eq:chi} holds if the \emph{factors} by which each feedback process suppresses the SFR are independent.  If, however, the feedback effects were additive and the \emph{amounts} by which they decrease the SFR were independent, then the equation would take on a slightly different form. The conclusion that the two feedback processes weaken each other also holds if they combine additively instead of mutliplicatively. This can easily be seen by considering that if SNe and AGN each reduce the SFR by more than a factor of 2, which is the case for much of the redshift and mass range, then their combined effect would yield a negative SFR if the two feedback processes were independent and combined additively.

We have verified that decreasing the simulation volume by a factor of eight, while keeping the resolution unchanged, has a negligible effect on the results.  If we decrease the numerical resolution then the magnitude of the feedback suppression becomes smaller, although the same qualitative trends hold as for the fiducial case, with SNe (AGN) dominating the suppression in low- (high-) mass objects, and an intermediate-mass regime where the two processes weaken one another's effects.  The halo mass at which AGN feedback begins to effectively suppress the SFR is higher for the low-resolution simulation because in this simulation gas densities at the centres of haloes are significantly lower than in the high-resolution case, so BHs grow more slowly and begin to affect galaxies at a later time \citep[see also][]{boot09}.  Thus, increasing the numerical resolution only strengthens our main conclusions.

\section*{Discussion}\label{sec:discussion}
Energetic feedback from SNe and from AGN are both important processes that are thought to control how much gas is able to condense into galaxies and form stars.  Using a set of cosmological SPH simulations that include, amongst other ingredients, both of these feedback processes, we have investigated how each (and both) of these processes affect the evolution of the contributions of haloes of fixed mass to the star formation history of the universe.  We demonstrated that, in the models, both AGN and SNe suppress the SFR when modelled in isolation, with SNe suppressing SF primarily in low-mass objects ($m_{200}<10^{12}\,\msun$) and AGN feedback operating in massive ($m_{200}>10^{12}\,\msun$) objects.  SNe are effective primarily in low-mass objects because, at high halo masses, hydrodynamic drag and radiative losses within the disc effectively confines the galactic winds.  AGN are effective primarily in high-mass objects because, in our prescription, seed mass black holes are only placed into massive, resolved haloes.  The exact halo mass range in which both feedback channels can operate is somewhat uncertain, but our results demonstrate that in any halo where both forms of feedback are important, their mutial interaction must be carefully considered.

However, if both AGN and SN feedback are included in a simulation, the factor by which the SFR is suppressed is significantly smaller than would be expected if the two processes had an independent effect on the SFR.  This occurs in the regime where both AGN and SN feedback are able to suppress the galaxy SFR coevally, i.e.\ in haloes of intermediate mass ($10^{11.25}\,\msun<m_{200}<10^{12}\,\msun$). The net effect is that they weaken one another's ability to suppress the galaxy SFR.

We caution the reader that our simulations have not fully converged with respect to resolution and that the halo mass at which BHs begin to affect the galaxy SFR changes with resolution.  Our simulations contain neither the numerical resolution or the physics to model the multi-phase interstellar medium.  The factor by which AGN and SN feedback suppress the SFR is therefore not well converged.  However, we verified that increasing the numerical resolution only strengthens the effects we discuss here. We note also that there is significant freedom in how we choose to model feedback processes in our simulations, and that taking significantly different choices about how and when energy is injected into galaxies may affect the results obtained.  

However, our finding that AGN and SN feedback weaken each other's effect on the SFR should not come as a surprise. Both SF and BH growth are thought to be self-regulated: feedback injects sufficient energy for the outflow to balance the accretion rate when averaged over sufficiently large length and time scales \citep[e.g.][]{boot10}. Since BH accretion episodes are generally accompanied by nuclear SF, both feedback processes will contribute to the nuclear outflow. If SN feedback is turned off, then the BH will compensate by injecting more energy. This extra AGN feedback will not only limit the BH's own growth, it will also reduce the SFR. Hence, we expect the reduction of the SFR due to AGN feedback to be greater in the absence of SN feedback.

Our work complements \citep{pawl09}, who found using similar methods that SN feedback and photo-heating by the UV background \emph{amplify} one another's effects on the SFR in low-mass galaxies.  SN feedback \lq puffs up\rq\, galaxies, making it easier for them to be evaporated by the UV background, and in turn the UV background removes gas from the outskirts of the galaxy, making it easier for SN feedback to drive out more gas.   The general conclusion that we can draw from these results, is that the way in which any one feedback process redistributes gas around the galaxy has the potential to either amplify or suppress the ability of other feedback processes to suppress the SFR. It is therefore vital that studies treat all feedback processes simultaneously and in a non-independent manner.

These results may have important implications for those semi-analytic models that make the implicit assumption that all types of feedback act independently of one another. However, as semi-analytic models do allow for indirect interactions between different feedback processes, it remains to be demonstrated whether these models can capture any of the effects discussed here. Finally, hydrodynamic simulations that model AGN in the absence of SN-driven winds \citep[e.g.][]{fabj10} may draw incorrect conclusions regarding the effect of AGN feedback.

\section*{Acknowledgments}
The authors would like to thank Andreas Pawlik and Alex Richings for a careful reading of the manuscript. The simulations were run on the Cosmology Machine at the Institute for Computational Cosmology in Durham (which is part of the DiRAC Facility jointly funded by STFC, the Large Facilities Capital Fund of BIS, and Durham University) as part of the Virgo Consortium research programme. This work benefited from support from the Netherlands Organisation for Scientific Research (NWO) and Marie Curie Initial Training Network CosmoComp (PITN-GA-2009-238356).

\section*{Author Contributions}
C.M.B. ran the simulations and participated in their analysis and interpretation and in writing the manuscript. J.S. participated in the analysis and interpretation of the simulations and in writing the manuscript.

\section*{Competing Interests}
The authors declare that they have no competing financial interests.

\end{document}